\documentclass[3p,times,twocolumn]{elsarticle_mod}

\usepackage{ecrc}

\volume{00}

\firstpage{1}

\journalname{Nuclear Physics B Proceedings Supplement}

\runauth{M.~Giordano and E.~Meggiolaro}

\jid{nuphbp}

\jnltitlelogo{Nuclear Physics B Proceedings Supplement}

\usepackage{amssymb}
\usepackage[figuresright]{rotating}

\newcommand{\la}{\langle}
\newcommand{\ra}{\rangle}

\newcommand{\f}[2]{\frac{#1}{#2}}

\begin{document}

\begin{frontmatter}

\dochead{}

\title{Hadronic total cross sections, Wilson loop correlators and the
  QCD spectrum} 

\author[l1]{Matteo Giordano\fnref{fn1}}
\ead{giordano@atomki.mta.hu}
\fntext[fn1]{Supported by the Hungarian Academy of Sciences under
  ``Lend\"ulet'' grant No. LP2011-011.}
\author[l2]{Enrico Meggiolaro}
\ead{enrico.meggiolaro@df.unipi.it}
\address[l1]{Institute for Nuclear Research of the Hungarian Academy of
  Sciences (ATOMKI), Bem t\'er 18/c, H--4026 Debrecen, Hungary}
\address[l2]{Dipartimento di Fisica, Universit\`a di Pisa,
and INFN, Sezione di Pisa, Largo Pontecorvo 3, I--56127 Pisa, Italy}

\begin{abstract}
We show how to obtain rising hadronic total cross sections in QCD, in
the framework of the nonperturbative approach to soft high-energy
scattering based on Wilson-loop correlators. Total cross sections turn
out to be of ``Froissart''-type, i.e., the leading energy dependence
is of the form $\sigma_{\rm tot} \sim B \log^2 s$, in agreement with
experiments. The observed universality of the prefactor $B$ is
obtained rather naturally in this framework. In this case, $B$ is
entirely determined by the stable spectrum of QCD, and predicted to be
$B_{th} = 0.22\, {\rm mb}$, in fair agreement with experiments.
\end{abstract}


\end{frontmatter}

\section{Introduction}
\label{sec:1}

Explaining the behaviour of hadronic total cross sections at high
energy is a very old problem, which is very rarely attacked within the
framework of QCD. Most of the approaches to this problem are based on
phenomenological models, which are sometimes QCD-inspired, but a full  
derivation from first principles of QCD is still lacking.

The observed rise of hadronic total cross sections at high energy is 
well described by a ``Froissart-like'' behaviour $\sigma_{tot} \sim B
\log^2 s$ with a universal prefactor $B\simeq 0.27\div 0.28\, {\rm
  mb}$~\cite{pdg}. This behaviour respects unitarity, as encoded in
the Froissart-\L ukaszuk-Martin bound~\cite{FLM1,FLM2,FLM3}, 
$ \sigma_{tot} \le B_{\rm FLM}\log^2 \f{s}{s_0}$, since $B\ll B_{\rm
  FLM}=\pi/m_\pi^2 \simeq 65\,{\rm mb}$.  

Besides deriving $\sigma_{tot}$ from the first principles of QCD, one
should be able to explain the observed universality of $B$, and also
the two orders of magnitude separating $B$ and $B_{\rm FLM}$. A step
in this direction has been made in Ref.~\cite{sigtot}, where we have
derived the asymptotic behaviour of hadronic total cross sections in
the framework of the nonperturbative approach to soft high-energy
scattering~\cite{Nac,Dos,pomeron-book}, finding indeed a
``Froissart-like'' behaviour, and a theoretical prediction for $B$ in 
fair agreement with experiments.

\section{Nonperturbative approach to soft  high-energy scattering}
\label{sec:2}

Understanding the rise of $\sigma_{tot}$ is part of the problem of
soft high-energy scattering, characterised by $|t|\le 1\, {\rm
  GeV}^2\ll s$. In this regime perturbation theory is not fully
reliable, and a nonperturbative approach is needed. In a nutshell, 
the nonperturbative approach to {\it elastic mes\-on-meson
  scattering} is as follows~\cite{Nac,Dos,pomeron-book}: 
\begin{enumerate}
\item mesons are described as wave packets of transverse colourless
  dipoles; 
\item in the soft high-energy regime, the dipoles travel essentially
  undisturbed on their classical, almost lightlike trajectories;
\item mesonic amplitudes are obtained from the dipole-dipole ($dd$)
  amplitudes after folding with the appropriate squared wave
  functions,   $A (s,b) =  \la\!\la A^{(dd)} (s, b;\nu) \ra\!\ra$.  
\end{enumerate}
Here $b$ is the impact parameter\footnote{We are interested in
  unpolarised scattering, so only the impact parameter modulus
  matters; for definiteness, we keep it oriented along direction 2 in
  the transverse plane.}, 
$\nu$ denotes collectively the dipole variables (longitudinal momentum
fraction, transverse size and orientation), and $\la\!\la \ldots
\ra\!\ra$ stands for integration over the dipole variables with the
mesonic wave functions. This approach extends also to processes
involving baryons, if one adopts for them a quark-diquark
picture~\cite{DR}. 

At high energy the $dd$ amplitude is given by the normalised connected
correlator ${\cal C}_M$ of the Wilson loops (WL) running along the
classical trajectories of the two dipoles, $A^{(dd)} (s,b;\nu)=-{\cal
  C}_M(\chi,{b};\nu)$, with $\chi \simeq \log\frac{s}{m^2}$ the
hyperbolic angle between the trajectories, and $m$ the mass of the
mesons (taken to be equal for simplicity).  

\begin{figure}
  \centering
  \includegraphics[height=17.0em]{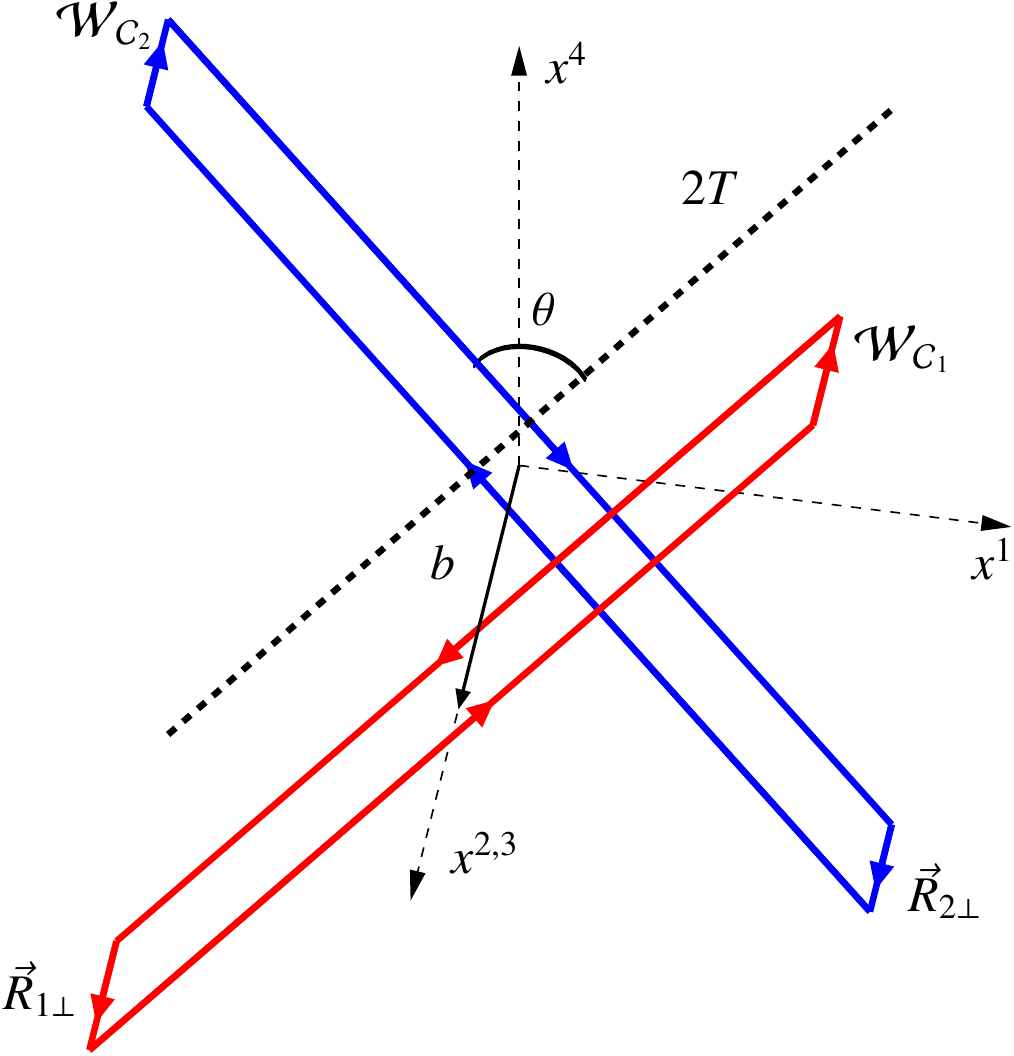}  
  \caption{The relevant Euclidean Wilson loops.}
  \label{fig:2}
\end{figure}
The correlator $\mathcal{C}_M$ is obtained from the correlator
$\mathcal{C}_E$ of two Euclidean WL at angle $\theta$ (see
Fig.~\ref{fig:2}), 
\begin{equation}
\mathcal{C}_E(\theta,{b};\nu) 
\equiv \lim_{T\to\infty}\frac{\langle \mathcal{W}_{\mathcal{C}_1}
  \mathcal{W}_{\mathcal{C}_2} \rangle}{\langle \mathcal{W}_{{\cal C}_1}\rangle 
  \langle \mathcal{W}_{{\cal C}_2} \rangle} - 1\,, 
\end{equation}
through the analytic continuation (AC)~\cite{Meggiolaro2005,EMduality} 
\begin{equation}
  \label{eq:ac}
  {\cal C}_M(\chi,b;\nu)= {\cal C}_E(\theta\to -i\chi,b;\nu)\,.
\end{equation}
A more detailed discussion of the approach can be found in
Ref.~\cite{sigtot} and references therein. The Euclidean formulation 
has allowed the study of the relevant WL correlator by means of
nonperturbative techniques, which include
instantons~\cite{ILM,GM2010}, the model of the stochastic 
vacuum~\cite{LLCM2}, holography~\cite{JP,JP2,Janik,GP2010,BKYZ}, and
the lattice~\cite{GM2010,GM2008,GMM}.

\section{Large-$s$ behaviour of $\sigma_{\rm tot}$ }
\label{sec:3}

The energy dependence of $\sigma_{\rm tot}$ is determined by the
large-$s$, large-$b$ behaviour of the amplitude through the
``effective radius'' of interaction $b_c=b_c(s)$, beyond which the
amplitude is negligible, as $\sigma_{\rm tot}\propto b_c^2$. 

To determine $b_c$, in Ref.~\cite{sigtot} we employed the following
strategy. On the Euclidean side, we obtain information on the $b$- and
$\theta$-dependencies of ${\cal C}_E$ by inserting between the WL
operators (for large enough $b$) a complete set of asymptotic states,
characterised by their particle content and by the momenta and spin of
each particle. After AC this gives us information on the $s$- and
$b$-dependencies. This requires two crucial analyticity assumptions:
\begin{enumerate}
\item AC can be performed separately for each term in the sum;
\item WL matrix elements are analytic in $\theta$.
\end{enumerate}
A few reasonable finiteness assumptions on the WL matrix elements are
also made. 

Under these assumptions, at large $\chi,b$, the relevant Minkowskian
correlator reads~\cite{sigtot} 
\begin{equation}
  \label{eq:correl}
  {\cal C}_M({\chi},{b};\nu)
  \simeq \sum_{\alpha\neq 0} 
  f_\alpha(\nu) \prod_{a} [{w_a}(\chi,b)]^{n_a(\alpha)}\,,
\end{equation}
where the sum is over (non-vacuum) states $\alpha$, and $n_a(\alpha)$
is the number of particles of type $a$ in state $\alpha$. Here
\begin{equation}
  \label{eq:ra}
  w_a(\chi,b) 
  =\f{e^{[r^{(a)}\chi-b] m^{(a)}}}{\sqrt{2\pi b m^{(a)}}}\,,
  \quad r^{(a)}\equiv \f{s^{(a)}-1}{m^{(a)}}\,,   
\end{equation}
with $(s^{(a)},m^{(a)})$ spin and mass of particles of type $a$, and
$f_\alpha$ are functions of the dipole variables only. Particles of
type $a$ contribute only for $b\lesssim r^{(a)}\chi$, and so the
effective radius of interaction is given by 
\begin{equation}
  \label{eq:bc}
  b_c(s)=\left(\max_a r^{(a)}\right) \log\f{s}{m^2} \equiv
  \f{1}{\mu}\log\f{s}{m^2}\,. 
\end{equation}
Here we assume the maximum to exist and to be positive. If it were
zero or negative, $\sigma_{\rm tot}$ would be constant or vanishing at
high energy. The spectrum is supposed to be free of massless states:
in case they were present and with spin at most 1, the maximisation
should be performed on the massive spectrum only~\cite{sigtot}. 

Using Eq.~\ref{eq:correl} we find for $\sigma_{\rm tot}$
\begin{equation}
  \label{eq:sigtotres}
   \sigma_{\rm tot} \mathop \simeq_{s\to\infty}
   2\pi(1-\kappa)[b_c(s)]^2
   \simeq
   \f{2\pi}{\mu^2}(1-\kappa)\log^2\!\f{s}{m^2} \,,
\end{equation}
with $|\kappa|\le 1$ due to unitarity~\cite{sigtot}. In general
$\kappa$ depends on the colliding hadrons. Analyticity and crossing 
symmetry~\cite{crossing,crossing2} requirements show that universality
is most naturally achieved if $\kappa=0$, corresponding to a vanishing
or oscillating correlator as $\chi\to\infty$ at fixed $b$.

\section{Total cross sections from the hadronic spectrum} 
\label{sec:6}

\begin{figure}
  \centering
  \includegraphics[height=13.2em]{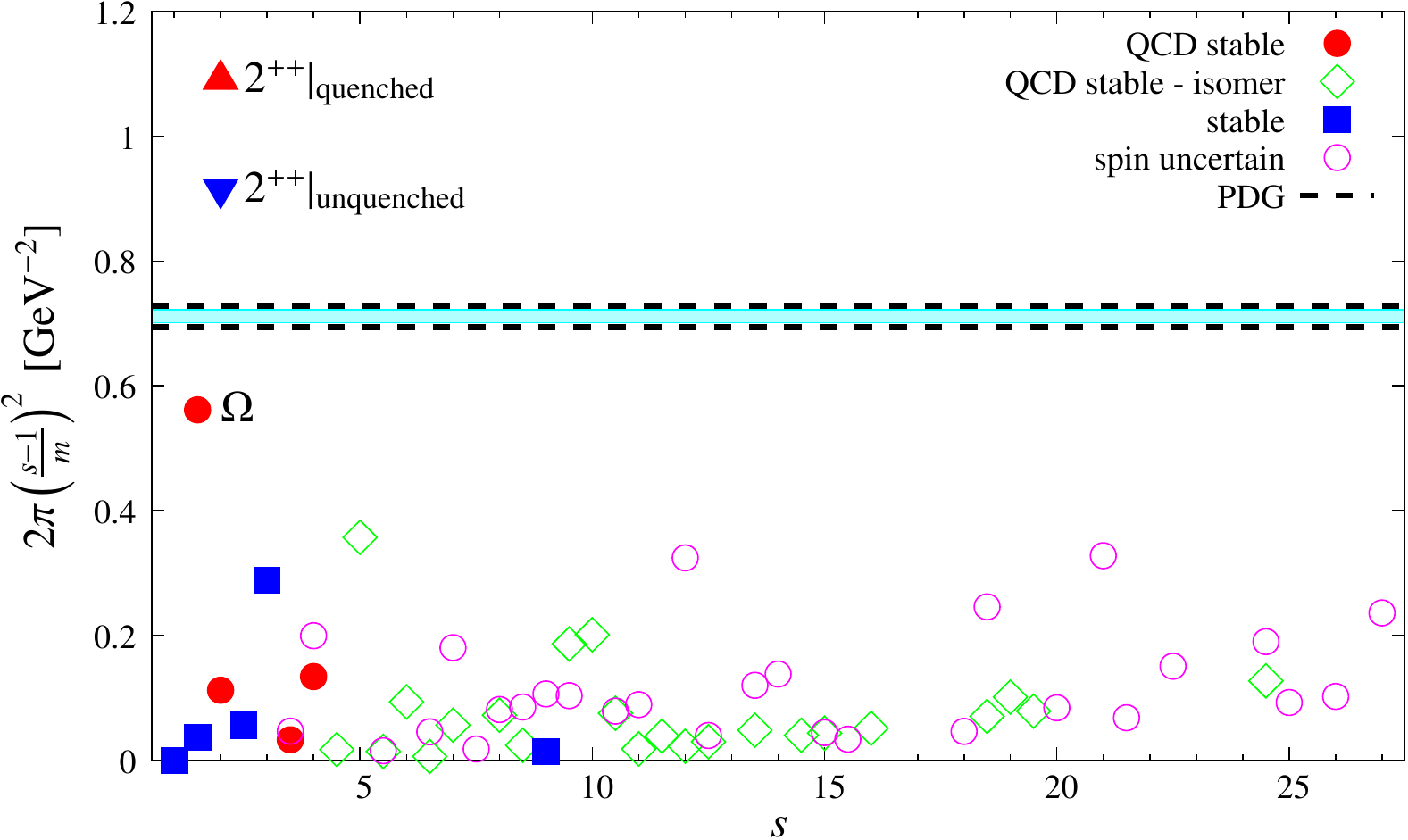}
  \caption{Plot of $2\pi\left(\f{s-1}{m}\right)^2$ for QCD-stable
    states. The $2^{++}$ glueball state~\cite{glueball,glueball2}
    and the experimental value for $B$~\cite{pdg} are also
    shown. Nuclear data are taken from Ref.~\cite{nubase}.} 
  \label{fig:3}
\end{figure}

The total cross section satisfies the bound~\cite{sigtot}
\begin{equation}
\label{eq:st1}
   \sigma_{\rm tot} 
\mathop \lesssim_{s\to\infty}
\f{4\pi}{\mu^2}\log^2\f{s}{m^2} = 2B_{\rm th}\log^2\f{s}{m^2}\,,  
\end{equation}
with $\mu^{-1} =\max_a r^{(a)}$ determined from the hadronic spectrum
by maximising $r^{(a)}$ over the stable states of QCD {\it in
  isolation}, as electroweak effects have been neglected from the
onset. Only states with $s^{(a)}\ge 1$ are considered, including
nuclei (see Fig.~\ref{fig:3}). 
 
Quite surprisingly, this singles out the $\Omega^\pm$ baryon, 
which yields $B_{\rm th}\simeq 0.56~{\rm GeV}^{-2}$. The resulting
bound on $\sigma_{\rm tot}$ is much more stringent than the original
bound, and the prefactor is of the same order of magnitude of the
experimental value $B_{\rm exp} \simeq 0.69\div 0.73~{\rm
  GeV}^{-2}$~\cite{pdg}. 
Furthermore, our bound is not singular in the chiral limit~\cite{largen}.

If universality is achieved as discussed above in Section \ref{sec:3}, 
then $\sigma_{\rm tot}$ is entirely determined by the hadronic spectrum,
and reads~\cite{sigtot} 
\begin{equation}
\label{eq:st2}
   \sigma_{\rm tot} \mathop \simeq_{s\to\infty}
   \f{2\pi}{\mu^2}\log^2\f{s}{m^2}=B_{\rm th}\log^2\f{s}{m^2}\,.
\end{equation}
This {\it prediction} for the prefactor is in fair agreement with
$B_{\rm exp}$, taking into account a systematic error of order $10\%$
on $B_{\rm exp}$, estimated by comparing the results of different
fitting procedures~\cite{pdg,Blogs,Blogs2,Blogs3}. The same conditions
leading to universal total cross sections also give universal,
black-disk-like elastic scattering amplitudes~\cite{sigtot}. 

Total cross sections are usually believed to be governed by the
gluonic sector of QCD. However, in the {\it quenched} theory one finds
from the glueball spectrum $B_Q\gtrsim 1.6 B_{\rm exp}$, suggesting
large unquenching effects~\cite{sigtot}.

\section{Conclusions}

In Ref.~\cite{sigtot} we have derived the asymptotic, high-energy
behaviour of hadronic total cross sections in the framework of the
nonperturbative approach to soft high-energy
scattering~\cite{Nac,Dos,pomeron-book}. We find a ``Froissart-like''
behaviour $\sigma_{tot}\sim B\log^2 s$, with $B$ (mainly) determined
by the hadronic spectrum (see Eqs.~\ref{eq:st1} and \ref{eq:st2}), and
in fair agreement with experiments.

Our main results do not depend on the detailed description of the
hadrons in terms of partons: adding gluons and sea quarks to the wave
functions would lead to more complicated WLs, but since our argument is
independent of their detailed form, our conclusions remain unchanged.

\nocite{*}
\bibliographystyle{elsarticle-num}
\bibliography{martin}

\end{document}